\documentclass[twocolumn,english,aps,pra,10pt,superscriptaddress,floatfix]{revtex4-2}
\usepackage{times}
\usepackage{graphicx}
\usepackage{bm}% bold math
\usepackage{amssymb,amsfonts,amsmath,amsbsy,bm,t1enc,latexsym}
\usepackage{float}
\usepackage[colorlinks=true,citecolor=blue,linkcolor=magenta]{hyperref}
\usepackage[markup=blue, authormarkupposition=left]{changes}
\usepackage[english]{babel}
\usepackage{url}
\usepackage{siunitx}
\DeclareSIUnit \dbc {dBc}
\usepackage{soul}
\usepackage[labelfont=bf, justification=centerlast]{caption}

\newcommand{\SiN}[0]{$\mathrm{Si}_3\mathrm{N}_4$}
\newcommand{\LN}[0]{$\mathrm{LiNbO}_3$} 
\newcommand{\LT}[0]{$\mathrm{LiTaO}_3$} 
 
\newcommand{\SiO}[0]{$\mathrm{SiO}_2$}

\begin{document}
\title{Lithium tantalate electro-optical photonic integrated circuits for high volume manufacturing}

\author{Chengli Wang}
\affiliation{National Key Laboratory of Materials for Integrated Circuits, Shanghai Institute of Microsystem and Information Technology, Chinese Academy of Sciences, 200050 Shanghai, China}
\affiliation{Institute of Physics, Swiss Federal Institute of Technology Lausanne (EPFL), CH-1015 Lausanne, Switzerland}
\affiliation{Center of Quantum Science and Engineering (EPFL), CH-1015 Lausanne, Switzerland}
\affiliation{These authors contributed equally to this work.}

\author{Zihan Li}
\affiliation{Institute of Physics, Swiss Federal Institute of Technology Lausanne (EPFL), CH-1015 Lausanne, Switzerland}
\affiliation{Center of Quantum Science and Engineering (EPFL), CH-1015 Lausanne, Switzerland}
\affiliation{These authors contributed equally to this work.}

\author{Johann Riemensberger}
\affiliation{Institute of Physics, Swiss Federal Institute of Technology Lausanne (EPFL), CH-1015 Lausanne, Switzerland}
\affiliation{Center of Quantum Science and Engineering (EPFL), CH-1015 Lausanne, Switzerland}

\author{Grigory Lihachev}
\affiliation{Institute of Physics, Swiss Federal Institute of Technology Lausanne (EPFL), CH-1015 Lausanne, Switzerland}
\affiliation{Center of Quantum Science and Engineering (EPFL), CH-1015 Lausanne, Switzerland}

\author{Mikhail Churaev}
\affiliation{Institute of Physics, Swiss Federal Institute of Technology Lausanne (EPFL), CH-1015 Lausanne, Switzerland}
\affiliation{Center of Quantum Science and Engineering (EPFL), CH-1015 Lausanne, Switzerland}

\author{Wil Kao}
\affiliation{Institute of Physics, Swiss Federal Institute of Technology Lausanne (EPFL), CH-1015 Lausanne, Switzerland}
\affiliation{Center of Quantum Science and Engineering (EPFL), CH-1015 Lausanne, Switzerland}

\author{Xinru Ji}
\affiliation{Institute of Physics, Swiss Federal Institute of Technology Lausanne (EPFL), CH-1015 Lausanne, Switzerland}
\affiliation{Center of Quantum Science and Engineering (EPFL), CH-1015 Lausanne, Switzerland}

\author{Terence Blesin}
\affiliation{Institute of Physics, Swiss Federal Institute of Technology Lausanne (EPFL), CH-1015 Lausanne, Switzerland}
\affiliation{Center of Quantum Science and Engineering (EPFL), CH-1015 Lausanne, Switzerland}

\author{Alisa Davydova}
\affiliation{Institute of Physics, Swiss Federal Institute of Technology Lausanne (EPFL), CH-1015 Lausanne, Switzerland}
\affiliation{Center of Quantum Science and Engineering (EPFL), CH-1015 Lausanne, Switzerland}

\author{Yang Chen}
\affiliation{National Key Laboratory of Materials for Integrated Circuits, Shanghai Institute of Microsystem and Information Technology, Chinese Academy of Sciences, 200050 Shanghai, China}

\author{Kai Huang}
\affiliation{National Key Laboratory of Materials for Integrated Circuits, Shanghai Institute of Microsystem and Information Technology, Chinese Academy of Sciences, 200050 Shanghai, China}

\author{Xi Wang}
\affiliation{National Key Laboratory of Materials for Integrated Circuits, Shanghai Institute of Microsystem and Information Technology, Chinese Academy of Sciences, 200050 Shanghai, China}

\author{Xin Ou}
\email[]{ouxin@mail.sim.ac.cn}
\affiliation{National Key Laboratory of Materials for Integrated Circuits, Shanghai Institute of Microsystem and Information Technology, Chinese Academy of Sciences, 200050 Shanghai, China}

\author{Tobias J. Kippenberg}
\email[]{tobias.kippenberg@epfl.ch}
\affiliation{Institute of Physics, Swiss Federal Institute of Technology Lausanne (EPFL), CH-1015 Lausanne, Switzerland}
\affiliation{Center of Quantum Science and Engineering (EPFL), CH-1015 Lausanne, Switzerland}

\maketitle

\subsection*{Abstract}
%\begin{abstract}
\textbf{
	Electro-optical photonic integrated circuits based on Lithium Niobate have demonstrated the vast capabilities afforded by material with a high Pockels coefficient \cite{zhu2021integrated,boes2023lithium}, allowing linear and high-speed modulators operating at CMOS voltage levels \cite{wang2018integrated} for applications ranging from data-center communications \cite{xu2022dual}, high-performance computing to photonic accelerators for AI \cite{shen2017deep}. 
	However despite major progress, the industrial adoption of this technology is compounded by the high cost per wafer, and limited wafer size – that result from the lack of existing high-volume applications in other domains, as is evidenced in the rapid adoption of silicon-on-insulator photonics driven by the vast investments into microelectronics.
	Here we overcome this challenge and demonstrate a photonic platform that satisfies the dichotomy of allowing scalable manufacturing at low cost – based on today’s existing infrastructure – while at the same time exhibiting equal, and superior properties to those of Lithium Niobate. 
	We demonstrate that it is possible to manufacture low loss photonic integrated circuits using Lithium Tantalate (\LT), a material that is already commercially adopted for acoustic filters in 5G and 6G \cite{butaud2020innovative}.
	We show that $\mathrm{LiTaO_3}$ posses equally attractive optical properties and can be etched with high precision and negligible residues using DUV lithography, diamond like carbon (DLC) as a hard mask \cite{li2022tightly} and alkaline wet etching.
	Using this approach we demonstrate microresonators with an intrinsic cavity linewidth of \SI{26.8}{\mega\hertz}, corresponding to a linear loss of \SI{5.6}{\decibel\per\meter} and demonstrate a Mach Zehnder modulator with $V_{\pi} L = 4.2$~V cm half-wave voltage length product.
	In comparison to Lithium Niobate, the photonic integrated circuits based on \LT~exhibit a much lower birefringence, allowing high-density circuits and broadband operation over all telecommunication bands (O to L band), exhibit higher photorefractive damage threshold, and lower microwave loss tangent.	Moreover, we show that the platform supports generation of soliton microcombs in X-Cut \LT~racetrack microresonator with electronically detectable repetition rate, i.e. \SI{30.1}{\giga\hertz}.
	Our work paves the way for scalable manufacturing of low-cost and large-volume high-performance electro-optical photonic integrated circuits for next-generation ultra high-speed modulators and devices for energy-efficient communication, computation or interconnects.
}

\section*{Introduction}
\begin{figure*}[t]
	\centering
	\includegraphics[width=1\linewidth]{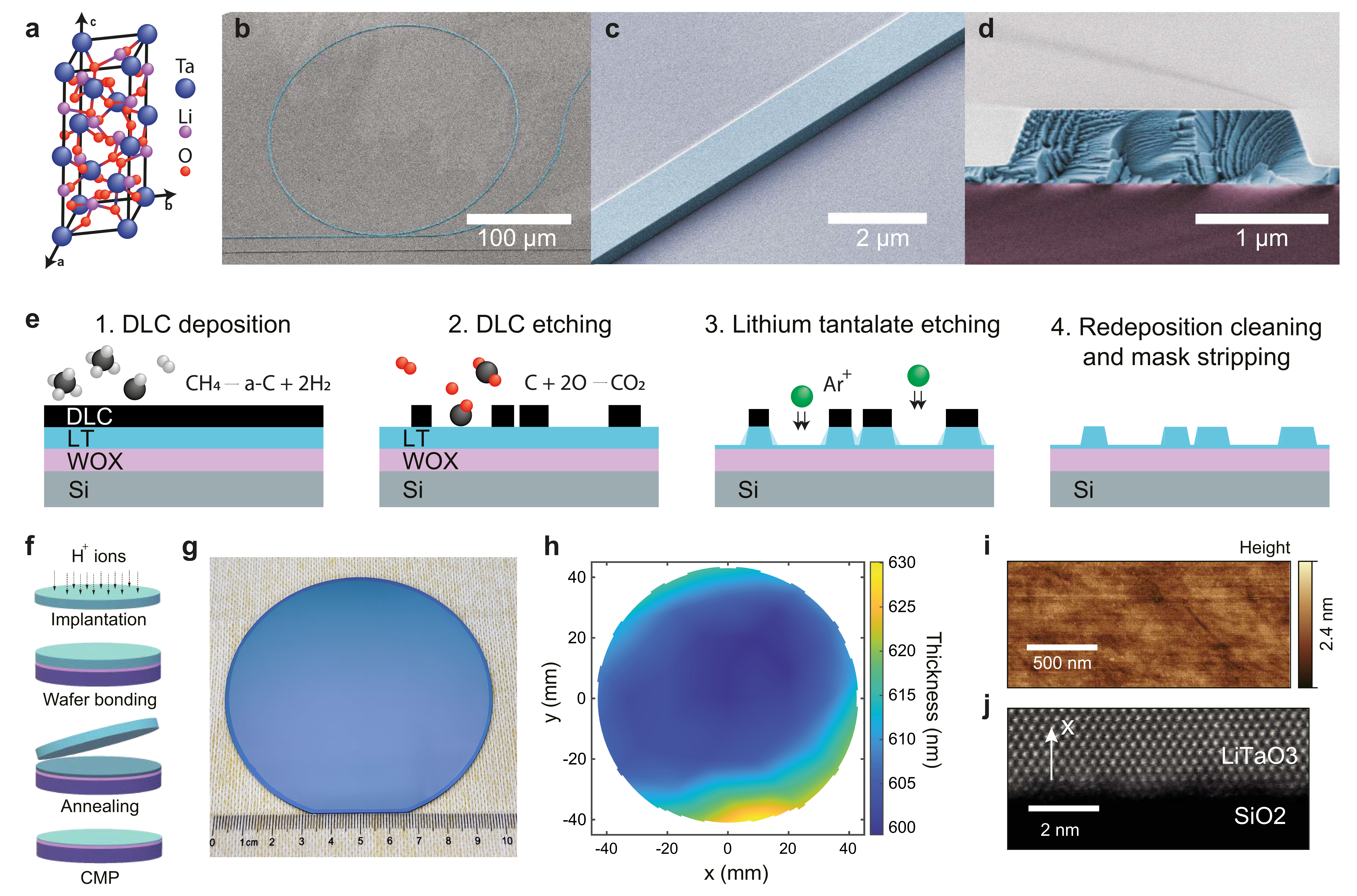}
	\caption{\textbf{Lithium-tantalate-on-insulator (LTOI) substrates and optical waveguides.} 
		(a) Crystallographic unit cell of \LT.
		(b) Colorized SEM of LTOI ring resonator.
		(c) Colorized scanning electron micrograph (SEM) of etched LTOI (blue) waveguide and sidewall.
		(d) Colorized SEM cross-section of etched LTOI waveguide on top of \SiO~bottom cladding (purple).
		(e) Schematic of fabrication workflow for LTOI optical waveguides including diamond-like carbon (DLC) hard mask deposition via plasma enhanced chemical vapor deposition (PECVD) from methane precursor, DLC dry etching via oxygen plasma, \LT~etching via argon ion beam etching (IBE), followed by redeposition and mask removal. \LT~is illustrated in blue, \SiO~in purple, DLC in black, and Si in grey.
		(f) Schematic of LTOI wafer bonding workflow with H-ion implantation, bonding, annealing, and CMP. 
		(g) Photograph of bonded wafer demonstrating uniform and defect-free bonding.
		(h) Thickness map of \LT~thin film on the wafer. 
		(i) Atomic force micrograph of the \LT~thin film surface.
		(j) High resolution STEM image of the \LT-\SiO~bonding interface.
	}\label{fig1}
\end{figure*}

\noindent Next-generation ultra-high-speed photonic integrated circuits (PICs) based on electro-optical materials are poised to play a role in energy-efficient data centers, optical communications, 5G/6G, or high-performance computing – granted that scalable low-cost manufacturing becomes possible. 
In the past two decades, photonic integrated circuits based on silicon (silicon photonics) have rapidly transitioned from academic research to widespread use in telecommunications \cite{thomson2016roadmap} and data centers \cite{xiang2021perspective}.
One crucial factor driving the commercial feasibility of this technological revolution is the high-volume availability and cost-effectiveness of silicon-on-insulator (SOI) wafers. 
These SOI wafers, prepared using the 'smart-cut', i.e. ion slicing techniques \cite{Bruel_1997}, enable the manufacturing of silicon photonic integrated circuits but crucially have a significantly larger magnitude of usage in consumer microelectronics. 
Today, globally more than 3 million wafers are produced per year in SOI, as large as 300 mm in diameter \cite{thomson2016roadmap}. Using the similar technique lithium niobate (\LN) has been fabricated into lithium niobate on insulator (LNOI) structures, offering an entirely new class of ultra high-speed, low voltage electro-optical photonic integrated circuits \cite{Wang2018,He2019,Zhang2019} that can become key components in future energy-efficient communication systems. 
Despite the tremendous scientific progress and the increased application range of \LN~photonic integrated circuits, the path to commercialization remains challenging, and commercial products do not exist to date. Unlike SOI technology, LNOI lacks a larger volume of consumer electronics driving its demand, resulting in economic limitations in its commercialization. 
In comparison, another ferroelectric material, lithium tantalate,  which shares similar structural properties with \LN, has entered a large-volume production stage driven by its applications in 5G filters \cite{ballandras2019new,yan2019wafer} and is projected to achieve a production capacity of 750k LTOI wafers per year by 2024 \cite{soitec2021}. 
The substantial volume enables significant benefits in terms of low-cost production when adopting LTOI as a platform for photonic integrated circuits - yet photonic integrated circuits based on this material have never been demonstrated to date.
\LT, in addition to the advantage in production volume, exhibits comparable or superior properties to \LN. \LT~is a traditional ferroelectric crystal with a nearly identical crystal structure as \LN~replacing Nb atoms in the crystal structure with heavier Ta atoms (cf. Figure~\ref{fig1}(a)).
The stronger chemical bonds induce a higher electron density in \LT~\cite{gruber2018atomistic}, which makes \LT~exhibit not only higher density but also increased stiffness and chemical stability. 
The optical bandgap of \LT~(3.93 eV) is larger than \LN~(3.63 eV) \cite{ccabuk1999urbach}, allowing e.g. nonlinear optical conversion to the visible and even ultraviolet \cite{meyn1997tunable} wavelength range,
with much decreased optical anisotropy, i.e. one order of magnitude reduced optical birefringence compared to \LN. The latter is particularly important as it allows manufacturing of tightly confining waveguides with strong bends without mode mixing, that can operate across all telecommunication bands simultaneously (from O to L band).
Moreover, \LT~features a comparable Pockels coefficient ($r_{33} = $ \SI{30.5}{\pico\meter\per\volt}) to the well-established \LN~with a moderately larger electrical permittivity $\epsilon_{33} = $ 43. 
In particular relevant for applications in the realm of microwave quantum transduction \cite{javerzac2016chip} the much lower microwave loss tangent of \LT~is a promising avenue to improve device performance to unity conversion efficiency, which has so far eluded efforts in \LN~due to limited quality factors of microwave resonators \cite{han2021microwave}.
Historically, despite the beneficial optical material properties, the use of \LT~for photonic devices in optical communication networks or scientific research has been limited. 
One of the reasons is that the Curie temperature of \LT~(\SI{610}{\degreeCelsius} - \SI{700}{\degreeCelsius}) is much lower than the temperatures needed for the fabrication of optical waveguides by ion diffusion (typically above \SI{1000}{\degreeCelsius}), which compounded the use of \LT~for bulk modulators based on ion diffused waveguide \cite{tormo2019low}. For this reason, legacy bulk modulator technology has employed \LN. 
Given the commercial adoption of LTOI in wireless applications due to its suitable acoustic properties, combined with the above optical properties, make it an ideal platform for scalable manufactured electro-optical photonic integrated circuits - yet the latter has to date never been demonstrated nor pursued.
Although free standing whispering gallery mode resonators have been fabricated from \LT~single crystals \cite{Soltani:16}, that utilize femtosecond laser direct writing \cite{doi.org/10.3390/molecules25173925} or focused ion beam milling \cite{yan2020high}, scalable manufactured photonic integrated circuits have remained an outstanding challenge.

\noindent Here, we overcome this challenge and implement the first photonic integrated circuit platform using \LT-on-insulator platform based on the direct etching \cite{li2022tightly} and demonstrate ultra-low optical loss, electro-optical tuning, switching via the Pockels effect, and soliton microcomb generation via the optical Kerr effect of \LT. 
We achieve this by transferring the DLC-based masking, etching process originally developed for \LN~to \LT~and proposing a new solution to remove \LT~redeposition, which highlights the flexibility of our process for the fabrication of a variety of ferroelectric photonics platforms. 
Equally, we demonstrate a DUV approach to electrode manufacturing. Taken together our work establishes a basis for scalable volume manufacturing of ultrahigh speed electro-optical photonic integrated circuits.

\section*{Low loss lithium tantalate-based photonic integrated circuits}  
\noindent The fabrication process for LTOI wafers and optical waveguides is depicted in Figure~\ref{fig1} and closely resembles recent efforts for LNOI \cite{li2022tightly}. 
We fabricated devices such as optical ring resonators (cf. Figure~\ref{fig1}(b)), racetrack resonators, and waveguide spirals (cf. Supplementary Figure~3).
Figures~\ref{fig1} (b,c) show the etched waveguide sidewalls and cleaved cross-section featuring low sidewall roughness and steep sidewall angles close to \SI{70}{\degree} with respect to the surface. 
The LTOI wafer was fabricated by the smart-cut technique \cite{yan2019wafer}.
The process flow is schematically illustrated in Figure~\ref{fig1}(f).
In contrast to the well-established LNOI preparation process that utilizes helium ion implantation, hydrogen ions are preferred for the fabrication of LTOI. The fabrication recipes of LTOI are more closely aligned with the high-volume commercial production of SOI wafers, resulting in higher efficiency and lower costs in the production of LTOI compared to LNOI.
We implanted the hydrogen ion into an X-Cut bulk \LT~ wafer with an energy of \SI{100}{\kilo\electronvolt} and a dose of $3.2\times 10^{16}$ /\si{\square\centi\meter}. 
After that, we flipped the implanted wafer and bonded it to a blank \SI{525}{\micro\meter} thick high resistivity silicon carrier wafer covered with \SI{2}{\micro\meter} thermal silicon dioxide followed by a thermal annealing step to separate the residual bulk wafer and the smart-cut \LT~thin film.
We perform edge removal of the \LT~thin film and chemical mechanical polishing to remove the rough and defect-rich layer of \LT~that was strongly affected by H-ion implantation and thin the \LT~film to the desired thickness of \SI{600}{\nano\meter}.
After chemical-mechanical polishing, the surface roughness is decreased to \SI{0.25}{\nano\meter} and the non-uniformity is less than \SI{30}{\nano\meter} over the full wafer and less than \SI{10}{\nano\meter} in the center area (cf. Figure~\ref{fig1}(h)). 
LTOI wafers with good uniformity and smooth surface roughness are the prerequisites for high-yield and low-loss PICs.
The crystallinity of \LT~and the \LT-\SiO~interface remain of high quality after the completion of the whole process, as well as the sharpness of the bonding interface, as is visible from the high-resolution scanning transmission electron microscopy (HRSTEM) image (cf. Figure~\ref{fig1}(j)). 

\noindent We fabricate \LT~photonic integrated circuits based on the process we recently demonstrated for deeply etched and tightly confining LNOI PICs \cite{li2022tightly}. 
First, we deposit a \SI{30}{\nano\meter} thick \SiN, \SI{480}{\nano\meter} thick layer of diamond-like-carbon (DLC) and \SI{60}{\nano\meter} thick \SiN~by plasma-enhanced chemical vapor deposition (PECVD) as the main hard mask for subsequent ion beam etching (IBE).
Then, we define the photonic waveguides and components by deep ultra-violet stepper photolithography and transfer the pattern first into a thin \SiN~layer by fluorine-based dry etching and subsequently into the DLC hard mask layer by oxygen-based dry etching in a reactive ion etcher.
The main etch of the photonic device layer is performed by IBE removing \SI{500}{\nano\meter} of \LT~and leaving a \SI{100}{\nano\meter} thick continuous \LT~slab across the wafer. 
We remove the \LT~redeposition on the waveguide sidewalls with an additional wet etching step and anneal the wafer at \SI{500}{\degreeCelsius} in an oxygen atmosphere. 
We deposit a \SI{2}{\micro\meter} thick upper cladding with PECVD based on a hydrogen-free precursor to avoid overtone absorption from optical phonons of the Si-OH stretch vibration around \SI{1.5}{\micro\meter}. 

\begin{figure*}[t]
	\centering
	\includegraphics[width=1\linewidth]{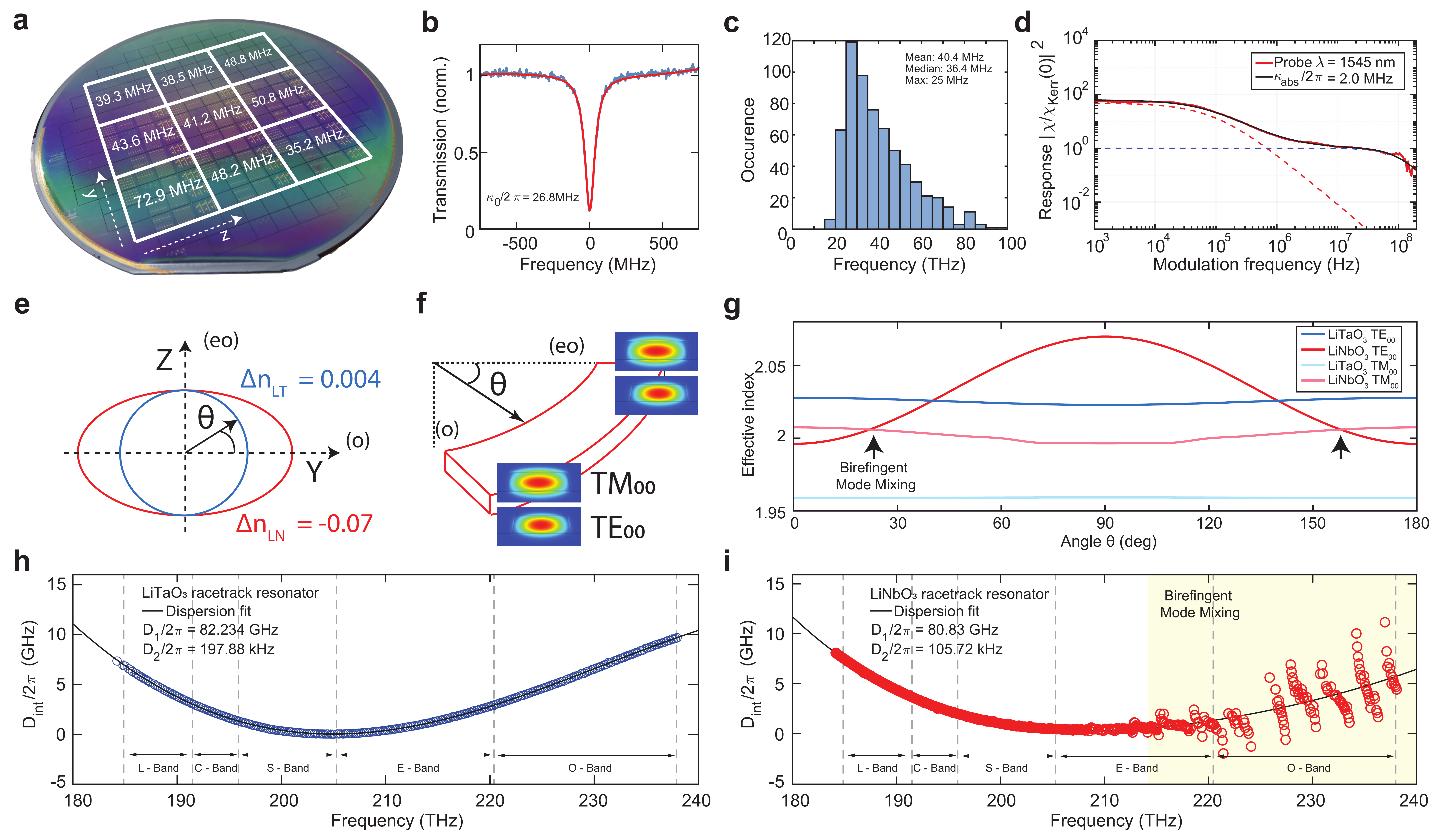}
	\caption{\textbf{Optical characterization of \LT~ photonic integrated circuits.}
		(a) Wafer-scale map of mean intrinsic loss $\kappa_0/2\pi$ for similar resonators fabricated using DUV stepper lithography.
		(b) Normalized resonance transmission spectrum of optical racetrack microresonator at 209.358~THz.
		(c) Statistical distribution of intrinsic loss $\kappa_0/2\pi$ of optical racetrack microresonator.
		(d) Nonlinear optical response measurement (solid red) and fit (solid black) of thermo-optical (red dashed) and Kerr (blue dashed) nonlinear response of optical microresonator demonstrating ultra-low optical absorption loss.
		(e) Illustration of \LN~ (red) strongly negative uniaxial and \LT~ (blue) weakly positive uniaxial crystal birefringence.
		(f) Illustration of curve angle and fundamental TE$_{00}$ and TM$_{00}$ mode profiles in LTOI.
		(g) Numerical simulation of fundamental TE$_{00}$ and TM$_{00}$ optical mode effective refractive indices of LNOI (blue) and LTOI (red) as a function of the angle between the waveguide and the Y-axis of the x-cut LN(T)OI film. The reduced birefringence of LTOI precludes unwanted birefringent mixing between fundamental TE$_{00}$/TM$_{00}$ modes in thick waveguides.
		(h) Dispersion profile of LTOI racetrack microresonator with cross-section \SI{2}{\micro\meter}~$\times$~\SI{0.5}{\micro\meter} and an \SI{100}{\nano\meter} thick slab. 
		(i) Dispersion profile of LNOI racetrack microresonator with similar cross-section and strong mode mixing at frequencies above \SI{215}{\tera\hertz}, which occupies the E-band and O-band in the optical communication. 		
	}\label{fig2}
\end{figure*}

\noindent Next, we optically characterize the \LT~PICs using frequency-comb calibrated tunable diode laser spectroscopy \cite{del2009frequency} to determine the optical loss and absorption of optical microresonators with waveguide width of \SI{2.0}{\micro\meter} across the 4-inch wafers (cf. Figure~\ref{fig2}(a)).
We find mean intrinsic loss rates $\kappa_0/2\pi$ between \SI{35.2}{\mega\hertz} and \SI{72.9}{\mega\hertz} with eight out of nine fields performing better than \SI{50.8}{\mega\hertz}.
The microresonator device loss of \SI{35.2}{\mega\hertz} corresponds to a propagation loss of $\alpha=$ \SI{7.3}{\decibel\per\meter} for the unreduced \LT~wafer that is specially used for optical applications. 
We also characterize the optical loss of the LTOI platform fabricated from the cheaper and more readily available acoustic grade \LT~bulk wafers. 
The acoustic LTOI exhibits slightly higher loss compared to the unreduced wafers.
The best field features a loss rate of \SI{42}{\mega\hertz} with a mean loss rate of \SI{82}{\mega\hertz} across the whole wafer (see Supplementary Figure 1).
This corresponds to losses of \SI{8.8}{\decibel\per\meter} and \SI{17.1}{\decibel\per\meter}, which is substantially below published losses in wafer-scale fabrication of LNOI PIC \cite{luke2020wafer}, making our process therefore directly applicable to widely used mass manufactured LTOI wafer substrates.
Figure~\ref{fig2}(b) depicts an optical resonance transmission spectrum and fit which indicates an intrinsic loss rate of \SI{26.8}{\mega\hertz}, which corresponds to a propagation loss of $\alpha=$ \SI{5.6}{\decibel\per\meter}.
We also fabricated optical waveguide spirals with a waveguide cross-section of \SI{1.75}{\micro\meter}$\times$\SI{0.6}{\micro\meter} and found a propagation loss around \SI{9}{\decibel\per\meter} (see Supplementary Figure 3).
Figure~\ref{fig3}(c) depicts the histogram of fitted intrinsic loss rates for the microresonator. 
The contributions of optical absorption and scattering from bulk and sidewall imperfections can be separated using thermal response spectroscopy \cite{liu2021high} (cf. Figure~\ref{fig2}(d)).
An intensity-modulated pump laser is tuned to the center of the optical resonance and the frequency modulation response of the optical microresonator due to the thermo-optical and Kerr effects is read out with a second laser tuned to the side of another resonance. 
We model the frequency dependence of the thermal effect due to the optical absorption and the optical Kerr effect using finite-element simulations and fit the combined response \cite{liu2021high,shams2022reduced}. 
We find that the absorption limit of our LTOI microresonator is \SI{2.0}{\mega\hertz}, corresponding to an absorption-limited propagation loss of \SI{0.4}{\decibel\per\meter} which is close to recent results obtained for LNOI \cite{shams2022reduced}. Therefore, the main source of loss in our tightly confining \LT~waveguides is dominated by scattering losses.

\noindent The optical birefringence of \LT~is more than one order of magnitude lower than in the case of \LN~(cf. Figure~\ref{fig2}(e)), which allows the fabrication of thick waveguides without incurring mode mixing between the fundamental modes in waveguide bends \cite{pan2019fundamental,li2022tightly}. 
Mode mixing occurs in x-cut \LN~waveguide bends when the TE mode transitions from the extraordinary (eo) to the ordinary (o) axes above a critical \LN~thickness that at wavelength \SI{1.55}{\micro\meter} lies around \SI{700}{\nano\meter} and for wavelength \SI{1.3}{\micro\meter} lies around \SI{600}{\nano\meter}, largely independent of the slab thickness, strongly constraining the design space for optical waveguides.
In contrast, the low and positive uniaxial birefringence of \LT~ precludes mode mixing in x-cut waveguides with a horizontal-to-vertical aspect ratio greater than one. 
We simulate the effective mode indices of the fundamental polarization modes of LNOI and LTOI for a waveguide thickness of \SI{600}{\nano\meter}, waveguide width of \SI{2}{\micro\meter} and wavelength of \SI{1.25}{\micro\meter} as a function of angle between the propagation and the eo crystal axes (cf. Figure~\ref{fig2}(g)).  
For \LN, we find a crossing of the fundamental TE and TM modes at an angle of \SI{25}{\degree}, while no mode crossing is found for an LTOI waveguide with the same dimension. 
This observation is in excellent agreement with the results from our optical dispersion measurement $D_\mathrm{int} = \omega_\mu - \omega_0 - D_1\cdot\mu$, where $\mu$ indicates the azimuthal mode index and $\omega_0/2\pi=205$ THz, for LTOI and LNOI waveguides, which are depicted in Figure~\ref{fig2}(h) and (i), respectively. 
Both optical microresonators have comparable anomalous dispersion, however, the dispersion profile of the LTOI microresonator remains smooth over the full measurement span from \SI{185}{\tera\hertz} to \SI{240}{\tera\hertz}, whereas the LNOI microresonator exhibits striking mode mixing at frequencies above \SI{215}{\tera\hertz}. Adjustments to the waveguide geometry and working wavelength can weaken the mode mixing caused by strong birefringence in LNOI \cite{wang2020polarization,li2022tightly}, however, such adjustments require sacrificing optical confinement and chip compactness. In comparison, LTOI offers much lower birefringence thereby providing greater flexibility in waveguide design and manufacturing and mode-mixing-free operation over all telecommunications bands from 1260 to 1625 nm, encompassing from O to L band.

\begin{figure*}[t]
	\centering
	\includegraphics[width=1\linewidth]{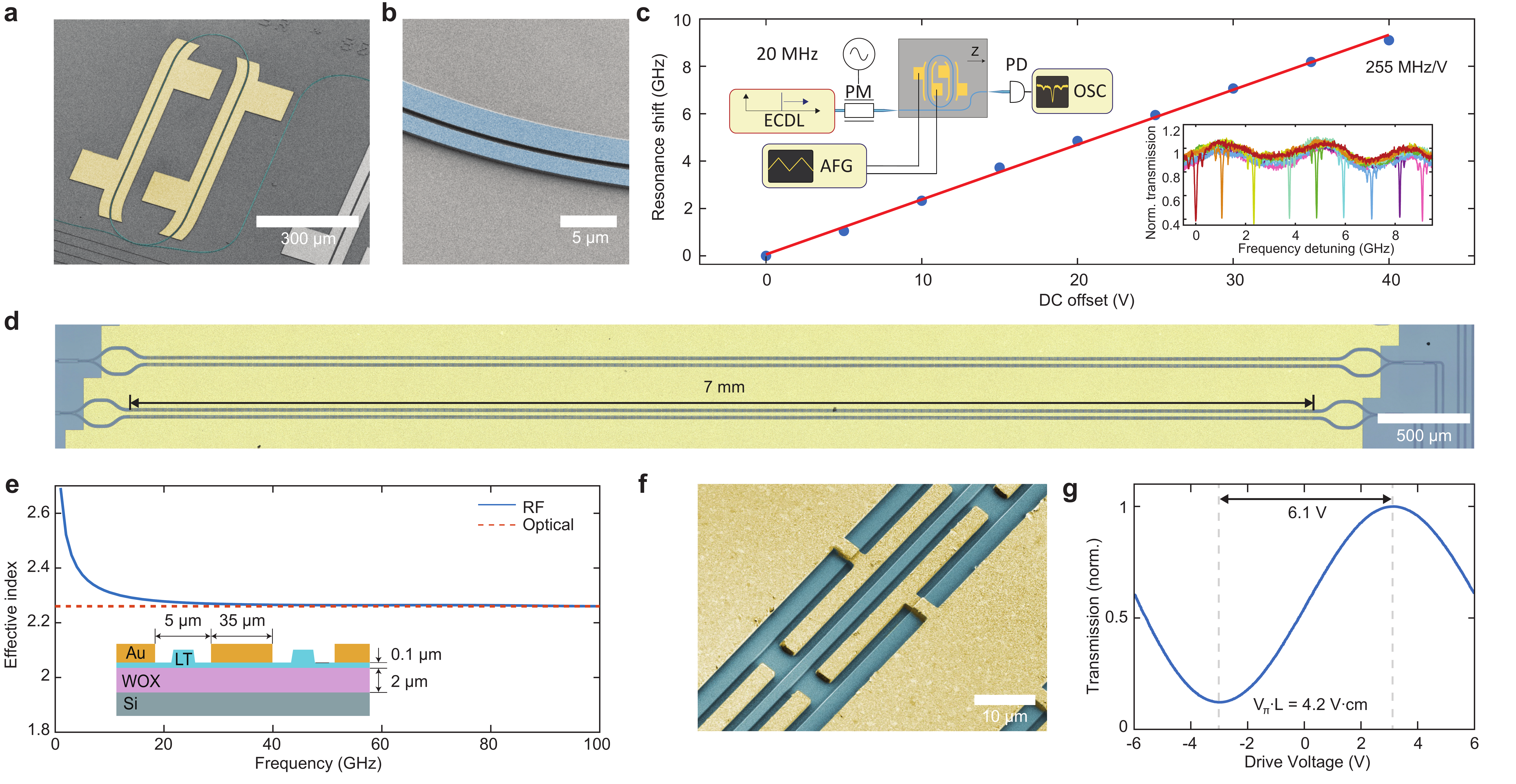}
	\caption{\textbf{Electro-optical tuning and switching in LTOI.} 
		(a) Colorized scanning electron micrograph (SEM) of LTOI (blue) racetrack optical microresonator with gold electrodes (yellow). 
		(b) Colorized SEM of pulley resonator and bus waveguide coupling section. 
		(c) Measured resonance shift as a result of tuning voltage. The linear fit indicates a voltage tuning response of 255~MHz/V. 
		Inset left: Schematic of measurement setup for microresonator tuning measurement with phase modulation sideband calibration. 
		Inset right: Electro-optical tuning of LTOI microresonator. Each color step corresponds to an increase in DC tuning voltage of 5~V. 
		(d) Optical micrograph of \SI{7}{\milli\meter} long Mach-Zehnder modulator (MZM).
		(e) Simulation of phase matching between the optical and microwave waves. Inset: schematic cross-section of waveguides and electrodes of the simulated structure.
		(f) Colorized SEM of MZM waveguides and electrodes.
		(g) Tuning curve of \SI{7}{\milli\meter} long MZM switch indicating a operation voltage of $V_\pi = 6.1$~V and voltage-length product $V_{\pi} L = 4.2$~V cm.
	}\label{fig3}
\end{figure*}

\section*{Electro-optical modulation}
\noindent To demonstrate the utility of the LTOI platform for electro-optics, we demonstrate a tunable high Q microresonator. 
The resonator has a racetrack design with apex radius \SI{100}{\micro\meter} and straight section length \SI{400}{\micro\meter} (cf. Figure~\ref{fig3}(a)) with a uniform waveguide width of \SI{2}{\micro\meter} and pulley-style coupling sections (cf. Figure~\ref{fig3}(b)). 
We applied a voltage across two of the four electrodes to measure the voltage tuning coefficient and measure the resonance position using an external cavity diode laser (ECDL) (cf. Figure~\ref{fig3}(c)).
We calibrated the laser frequency with a \SI{250}{\mega\hertz} phase modulation by detecting the sidebands around the resonance. 
We find a voltage tuning efficiency of \SI{255}{\mega\hertz}/V using a single electrode pair, which corresponds to \SI{510}{\mega\hertz}/V if both phase shifter sections are modulated.  
We also fabricated a 2$\times$2 electro-optical switch based on a Mach Zehnder interferometer (MZI) composed of two 2$\times$2 multimode interference (MMI) beam splitters at either end and a push-pull optical waveguide phase shifter pair with a length of \SI{7}{\milli\meter}((cf. Figure~\ref{fig3}(d))).
The waveguide width is \SI{1.5}{\micro\meter} and the gap between the \LT~waveguide sidewalls and the gold electrode is \SI{2}{\micro\meter} on each side.
Metal electrodes are fabricated using a DUV-lithography-based lift-off process that allows us to manufacture electrodes with a thickness of \SI{800}{\nano\meter} and with an alignment tolerance below \SI{100}{\nano\meter} to the optical waveguide.
The transmission through the switch is plotted in Figure~\ref{fig3}(g). 
The switching contrast is \SI{10}{\decibel}, which is commensurate with an imbalance of 5\% for the 2$\times$2 MMI beam splitters.
We further find $V_\pi = $ \SI{6.1}{\volt} (cf. Figure~\ref{fig3}(g)), which corresponds a voltage modulation efficiency of $V_\pi L = $ \SI{4.2}{\volt\centi\meter} for the push-pull MZI. The larger dielectric constant of \LT~($\epsilon_{LT} = $ 43) is generally considered unfavourable for phase matching between optical and microwave fields \cite{wooten2000review}. However, in our LTOI platform, as the optical fields is tightly localized in the sub-micrometre waveguide region while the electrical field primarily resides in the low dielectric silica layer ($\epsilon_{silica} = $ 3.9), it is convenience to engineer the RF and optical group velocities independently. Simulation results demonstrate that a conventional design can maintain velocity matching between the optical and the microwave signals at very high microwave frequencies without compromising the electro-optic efficiency (cf. Figure~\ref{fig3}(e)). By combining the demonstrated low propagation loss and high electro-optical efficiency in our LTOI platform with a phase-matched electrode transmission line could lead to a competitive bandwidth comparable to the well-researched LNOI platform. 

\begin{figure*}[t]
	\centering
	\includegraphics[width=1\linewidth]{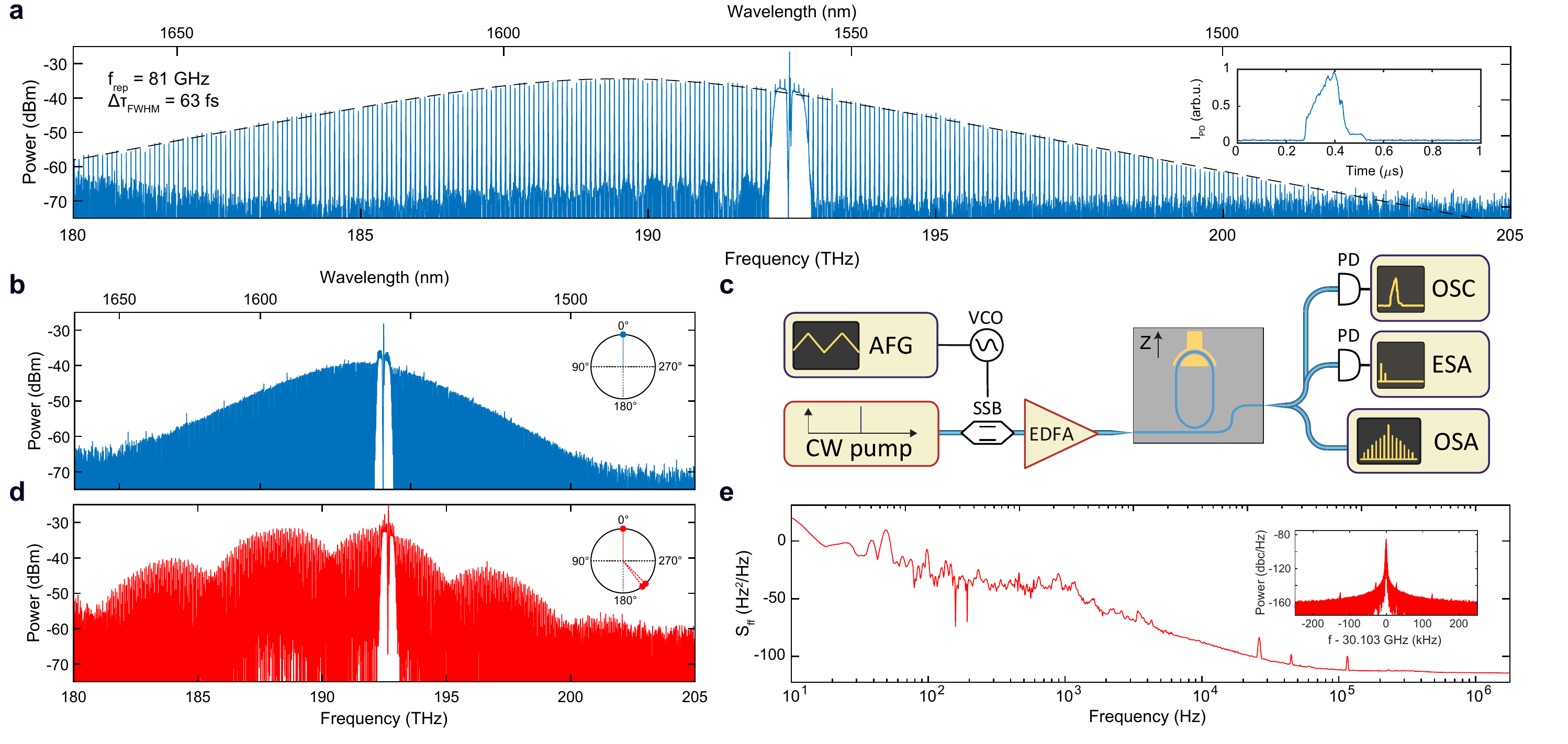}
	\caption{\textbf{Dissipative Kerr soliton (DKS) generation in \LT~ photonic integrated circuit based microresonators} 
		(a) Optical spectrum of a single soliton microcomb featuring a $sech^2$-spectral profile with a 3~dB bandwidth of 4.9~THz corresponding to an FWHM pulse duration of 63~fs at a pulse repetition rate $f_\mathrm{rep}$ of 81~GHz. The inset shows the generated light during the rapid laser scan measured by filtering out the pump light. 
		(b) Optical spectrum of a single soliton with repetition rate 30.1~GHz. 
		(c) Optical setup for soliton generation in LTOI. Rapid laser scans are generated using a single-sideband modulator (SSB) and an erbium-doped fiber amplifier. The soliton microcombs are analyzed using an optical spectrum analyzer (OSA) and the nonlinearly generated light and microwave beat notes are recorded with a fast photodiode (PD) and analyzed with an oscilloscope (OSC) and electrical spectrum analyzer (ESA), respectively.
		(d) Optical spectrum of three-soliton state with repetition rate 30.1~GHz. 
		(e) Single-side band phase noise power spectral density of 30.1~GHz microwave beat note generated from the multi-soliton state of the panel (d). Inset: Spectrum of microwave beat note with resolution bandwidth 30~Hz. 
	}\label{fig4}
\end{figure*}

\section*{Soliton microcomb generation}
\noindent Finally, we investigate the \LT~platform for nonlinear microcomb generation. The strong optical confinement, high Q-factor, anomalous dispersion, and substantial Kerr nonlinearity of our LTOI microresonators make them naturally suitable for dissipative Kerr soliton (DKS) generation \cite{herr2014temporal,kippenberg2018dissipative}. 
We achieved single soliton generation at pulse repetition rates \SI{81}{\giga\hertz} (cf. Figure ~\ref{fig4}(a)) and \SI{30.1}{\giga\hertz} (cf. Figure ~\ref{fig4}(b)). 
The optical setup for single soliton generation is depicted in Figure ~\ref{fig4}(c). 
We utilized a rapid single sideband tuning scheme pioneered in ref. \cite{stone2018thermal} to overcome thermal nonlinearities and initiate solitons at a pump power of 90~mW on-chip using an external cavity diode laser (ECDL) and an erbium-doped fiber amplifier for pumping. 
The FWHM spectral bandwidth of the \SI{81}{\giga\hertz} single soliton is \SI{4.9}{\tera\hertz}, corresponding to a compressed pulse duration of \SI{63}{\femto\second}. 
The 30.1~GHz single soliton state features a bandwidth of \SI{4.0}{\tera\hertz} and supports a pulse duration of \SI{71}{\femto\second}.
Various multisoliton states were also achieved and we depict an example state with three solitons in Figure~\ref{fig4}(d). 
The low repetition rate of our \SI{30.1}{\giga\hertz} solitons allows the direct detection of the microwave repetition beat note on a fast photodiode. 
We measure the phase noise of the microwave beat note using an electrical spectrum analyzer (ESA) and find a phase noise level of \SI{-86}{\dbc\per\hertz} at an offset frequency of \SI{10}{\kilo\hertz} and \SI{-114}{\dbc\per\hertz} at an offset frequency of \SI{1}{\mega\hertz}, higher than earlier measurements using \SiN~optical microresonators \cite{liu2020photonic} and in z-cut \LN~ \cite{He2023}. 
It is notable that here DKS generation was achieved in an x-cut ferroelectric crystal sample for the first time. 

\section*{Conclusion}
\noindent In summary, we have developed lithium tantalate photonic integrated circuits that are low loss, exhibit low birefringence and have comparable properties to Lithium Niobate, but crucially employ a material that is already used today commercially in large volumes for wireless filters, thereby providing a path to scalable manufacturing at low cost. LTOI PICs achieve comparable loss and electro-optical performance to well-established LNOI that have major potential for penetrating datacenter interconnects \cite{wang2018integrated}, long-haul optical communications \cite{xu2022dual}, and quantum photonics \cite{zhao2020high,nehra2022few} to name just a few applications.
The use of low cost substrates is of central importance for adoption in applications such as data center interconnects, where the die size is large due to the requirements of low modulator voltage and the length of travelling wave modulator devices.
In our work, we do not only establish a smart-cut process for the manufacturing of LTOI wafer substrates, but also demonstrate a complete manufacturing process including the etching of \LT, the removal of the redeposition of etch products on the waveguide sidewall, and the manufacturing of thick metal electrodes for functional electro-optic devices and demonstrate key performance metrics such as low propagation losses of \SI{5.6}{\decibel\per\meter}.   
Our process is fully wafer-scale and based on deep ultraviolet photolithography and lays the foundation to scalable manufacturing of high performance electro-optical PICs that can harness the scale of LTOI wafer fabrication for RF filters, which is ongoing both on 150~mm and 200~mm wafer size.
Our LTOI platform is, in particular, promising for applications that can directly exploit the superior properties of the material such as the reduced birefringence, where our platform is capable of processing signals across all optical communications bands (1260-1620~nm) in a single PIC due to the successful suppression of fundamental mode mixing and does support soliton microcomb generation also in the X-Cut, in contrast to \LN~ where soliton microcomb generation has only been observed in Z-Cut so far \cite{he2019self,He2023}, which has compounded the combination of electro-optical and Kerr nonlinearities \cite{gong2022monolithic}.
LTOI is also equally promising for quantum transduction of single microwave photons \cite{javerzac2016chip,han2021microwave} that has recently garnered attention to overcome the thermal bottlenecks of interfacing with superconducting quantum computers \cite{youssefi2021cryogenic}, because of the fact that the dielectric loss tangent of \LT~\cite{jacob2004temperature} is much lower than \LN~\cite{yang2007characteristics}.

\section*{Author Contributions}
C.W., Y.C., and K.H. fabricated the LTOI wafers. 
Z.L. fabricated the LTOI PICs. 
J.R. and Z.L. designed the PICs.
C.W. and Z.L. characterized the samples. 
C.W., Z.L., M.C., X.J., G.L., T.B., and W.K. performed optical experiments.
J.R., C.W, Z.L. analyzed the data, prepared the figures and wrote the manuscript with input from all authors.  
T.J.K. and X.Ou supervised the project.  

\section*{Funding Information}

T.J.K. acknowledges funding from the European Research Council grant no. 835329 (ExCOM-cCEO) and from the EU Horizon Europe research and innovation program through grant no. 101113260 (HDLN). J.R. acknowledges funding from the SNSF through an Ambizione Fellowship no. 201923. C.W. acknowledges financial support from China Scholarship Council (No.202104910464). X.O. acknowledges the National Key R\&D Program (No.2022YFA-1404601) from the Ministry of Science and Technology of China. 

\section*{Data Availability}
The code, data, and micrographs used to produce the plots within this work will be released on the repository \texttt{Zenodo} upon publication of this preprint.

\section*{Acknowledgements}
The samples were fabricated in the EPFL Center of MicroNanoTechnology (CMi) and the Institute of Physics (IPHYS) cleanroom. The LTOI wafers were fabricated in Shanghai Novel Si Integration Technology (NSIT) Co., Ltd. and the SIMIT-CAS.

\section*{Competing interests}
The authors declare no competing financial interests.

\bibliographystyle{naturemag}

\begin{thebibliography}{10}
	\expandafter\ifx\csname url\endcsname\relax
	\def\url#1{\texttt{#1}}\fi
	\expandafter\ifx\csname urlprefix\endcsname\relax\def\urlprefix{URL }\fi
	\providecommand{\bibinfo}[2]{#2}
	\providecommand{\eprint}[2][]{\url{#2}}
	
	\bibitem{zhu2021integrated}
	\bibinfo{author}{Zhu, D.} \emph{et~al.}
	\newblock \bibinfo{title}{Integrated photonics on thin-film lithium niobate}.
	\newblock \emph{\bibinfo{journal}{Advances in Optics and Photonics}}
	\textbf{\bibinfo{volume}{13}}, \bibinfo{pages}{242--352}
	(\bibinfo{year}{2021}).
	
	\bibitem{boes2023lithium}
	\bibinfo{author}{Boes, A.} \emph{et~al.}
	\newblock \bibinfo{title}{Lithium niobate photonics: Unlocking the
		electromagnetic spectrum}.
	\newblock \emph{\bibinfo{journal}{Science}} \textbf{\bibinfo{volume}{379}},
	\bibinfo{pages}{eabj4396} (\bibinfo{year}{2023}).
	
	\bibitem{wang2018integrated}
	\bibinfo{author}{Wang, C.} \emph{et~al.}
	\newblock \bibinfo{title}{Integrated lithium niobate electro-optic modulators
		operating at {CMOS}-compatible voltages}.
	\newblock \emph{\bibinfo{journal}{Nature}} \textbf{\bibinfo{volume}{562}},
	\bibinfo{pages}{101--104} (\bibinfo{year}{2018}).
	
	\bibitem{xu2022dual}
	\bibinfo{author}{Xu, M.} \emph{et~al.}
	\newblock \bibinfo{title}{Dual-polarization thin-film lithium niobate in-phase
		quadrature modulators for terabit-per-second transmission}.
	\newblock \emph{\bibinfo{journal}{Optica}} \textbf{\bibinfo{volume}{9}},
	\bibinfo{pages}{61--62} (\bibinfo{year}{2022}).
	
	\bibitem{shen2017deep}
	\bibinfo{author}{Shen, Y.} \emph{et~al.}
	\newblock \bibinfo{title}{Deep learning with coherent nanophotonic circuits}.
	\newblock \emph{\bibinfo{journal}{Nature photonics}}
	\textbf{\bibinfo{volume}{11}}, \bibinfo{pages}{441--446}
	(\bibinfo{year}{2017}).
	
	\bibitem{butaud2020innovative}
	\bibinfo{author}{Butaud, E.} \emph{et~al.}
	\newblock \bibinfo{title}{Innovative smart cut™ piezo on insulator (poi)
		substrates for 5g acoustic filters}.
	\newblock In \emph{\bibinfo{booktitle}{2020 IEEE International Electron Devices
			Meeting (IEDM)}}, \bibinfo{pages}{34--6} (\bibinfo{organization}{IEEE},
	\bibinfo{year}{2020}).
	
	\bibitem{li2022tightly}
	\bibinfo{author}{Li, Z.} \emph{et~al.}
	\newblock \bibinfo{title}{Tightly confining lithium niobate photonic integrated
		circuits and lasers}.
	\newblock \emph{\bibinfo{journal}{arXiv preprint arXiv:2208.05556}}
	(\bibinfo{year}{2022}).
	
	\bibitem{thomson2016roadmap}
	\bibinfo{author}{Thomson, D.} \emph{et~al.}
	\newblock \bibinfo{title}{Roadmap on silicon photonics}.
	\newblock \emph{\bibinfo{journal}{Journal of Optics}}
	\textbf{\bibinfo{volume}{18}}, \bibinfo{pages}{073003}
	(\bibinfo{year}{2016}).
	
	\bibitem{xiang2021perspective}
	\bibinfo{author}{Xiang, C.}, \bibinfo{author}{Bowers, S.~M.},
	\bibinfo{author}{Bjorlin, A.}, \bibinfo{author}{Blum, R.} \&
	\bibinfo{author}{Bowers, J.~E.}
	\newblock \bibinfo{title}{Perspective on the future of silicon photonics and
		electronics}.
	\newblock \emph{\bibinfo{journal}{Applied Physics Letters}}
	\textbf{\bibinfo{volume}{118}} (\bibinfo{year}{2021}).
	
	\bibitem{Bruel_1997}
	\bibinfo{author}{Bruel, M.} \& \bibinfo{author}{Auberton-Hervé, B.~A.}
	\newblock \bibinfo{title}{Smart-cut: A new silicon on insulator material
		technology based on hydrogen implantation and wafer bonding}.
	\newblock \emph{\bibinfo{journal}{Japanese Journal of Applied Physics}}
	\textbf{\bibinfo{volume}{36}}, \bibinfo{pages}{1636} (\bibinfo{year}{1997}).
	
	\bibitem{Wang2018}
	\bibinfo{author}{Wang, C.} \emph{et~al.}
	\newblock \bibinfo{title}{{Integrated lithium niobate electro-optic modulators
			operating at CMOS-compatible voltages}}.
	\newblock \emph{\bibinfo{journal}{Nature}} \textbf{\bibinfo{volume}{562}},
	\bibinfo{pages}{101--104} (\bibinfo{year}{2018}).
	
	\bibitem{He2019}
	\bibinfo{author}{He, M.} \emph{et~al.}
	\newblock \bibinfo{title}{High-performance hybrid silicon and lithium niobate
		mach--zehnder modulators for 100 gbit{\thinspace}s-1 and beyond}.
	\newblock \emph{\bibinfo{journal}{Nature Photonics}}
	\textbf{\bibinfo{volume}{13}}, \bibinfo{pages}{359--364}
	(\bibinfo{year}{2019}).
	
	\bibitem{Zhang2019}
	\bibinfo{author}{Zhang, M.} \emph{et~al.}
	\newblock \bibinfo{title}{Broadband electro-optic frequency comb generation in
		a lithium niobate microring resonator}.
	\newblock \emph{\bibinfo{journal}{Nature}} \textbf{\bibinfo{volume}{568}},
	\bibinfo{pages}{373--377} (\bibinfo{year}{2019}).
	
	\bibitem{ballandras2019new}
	\bibinfo{author}{Ballandras, S.} \emph{et~al.}
	\newblock \bibinfo{title}{New generation of saw devices on advanced engineered
		substrates combining piezoelectric single crystals and silicon}.
	\newblock In \emph{\bibinfo{booktitle}{2019 Joint Conference of the IEEE
			International Frequency Control Symposium and European Frequency and Time
			Forum (EFTF/IFC)}}, \bibinfo{pages}{1--6} (\bibinfo{organization}{IEEE},
	\bibinfo{year}{2019}).
	
	\bibitem{yan2019wafer}
	\bibinfo{author}{Yan, Y.} \emph{et~al.}
	\newblock \bibinfo{title}{Wafer-scale fabrication of 42° rotated y-cut
		litao3-on-insulator (ltoi) substrate for a saw resonator}.
	\newblock \emph{\bibinfo{journal}{ACS Applied Electronic Materials}}
	\textbf{\bibinfo{volume}{1}}, \bibinfo{pages}{1660--1666}
	(\bibinfo{year}{2019}).
	
	\bibitem{soitec2021}
	\bibinfo{author}{SOITEC}.
	\newblock \bibinfo{title}{Capital markets day 2021} (\bibinfo{year}{2021}).
	\newblock
	\bibinfo{note}{\url{https://www.soitec.com/en/capital-markets-day-2021}}.
	
	\bibitem{gruber2018atomistic}
	\bibinfo{author}{Gruber, M.} \emph{et~al.}
	\newblock \bibinfo{title}{Atomistic origins of the differences in anisotropic
		fracture behaviour of litao3 and linbo3 single crystals}.
	\newblock \emph{\bibinfo{journal}{Acta Materialia}}
	\textbf{\bibinfo{volume}{150}}, \bibinfo{pages}{373--380}
	(\bibinfo{year}{2018}).
	
	\bibitem{ccabuk1999urbach}
	\bibinfo{author}{{\c{C}}abuk, S.} \& \bibinfo{author}{Mamedov, A.}
	\newblock \bibinfo{title}{Urbach rule and optical properties of the linbo3 and
		litao3}.
	\newblock \emph{\bibinfo{journal}{Journal of Optics A: Pure and Applied
			Optics}} \textbf{\bibinfo{volume}{1}}, \bibinfo{pages}{424}
	(\bibinfo{year}{1999}).
	
	\bibitem{meyn1997tunable}
	\bibinfo{author}{Meyn, J.-P.} \& \bibinfo{author}{Fejer, M.}
	\newblock \bibinfo{title}{Tunable ultraviolet radiation by second-harmonic
		generation in periodically poled lithium tantalate}.
	\newblock \emph{\bibinfo{journal}{Optics letters}}
	\textbf{\bibinfo{volume}{22}}, \bibinfo{pages}{1214--1216}
	(\bibinfo{year}{1997}).
	
	\bibitem{javerzac2016chip}
	\bibinfo{author}{Javerzac-Galy, C.} \emph{et~al.}
	\newblock \bibinfo{title}{On-chip microwave-to-optical quantum coherent
		converter based on a superconducting resonator coupled to an electro-optic
		microresonator}.
	\newblock \emph{\bibinfo{journal}{Physical Review A}}
	\textbf{\bibinfo{volume}{94}}, \bibinfo{pages}{053815}
	(\bibinfo{year}{2016}).
	
	\bibitem{han2021microwave}
	\bibinfo{author}{Han, X.}, \bibinfo{author}{Fu, W.}, \bibinfo{author}{Zou,
		C.-L.}, \bibinfo{author}{Jiang, L.} \& \bibinfo{author}{Tang, H.~X.}
	\newblock \bibinfo{title}{Microwave-optical quantum frequency conversion}.
	\newblock \emph{\bibinfo{journal}{Optica}} \textbf{\bibinfo{volume}{8}},
	\bibinfo{pages}{1050--1064} (\bibinfo{year}{2021}).
	
	\bibitem{tormo2019low}
	\bibinfo{author}{Tormo-Marquez, V.}, \bibinfo{author}{D{\'\i}az-Hijar, M.},
	\bibinfo{author}{Carrascosa, M.}, \bibinfo{author}{Shur, V.~Y.} \&
	\bibinfo{author}{Olivares, J.}
	\newblock \bibinfo{title}{Low loss optical waveguides fabricated in litao 3 by
		swift heavy ion irradiation}.
	\newblock \emph{\bibinfo{journal}{Optics Express}}
	\textbf{\bibinfo{volume}{27}}, \bibinfo{pages}{8696--8708}
	(\bibinfo{year}{2019}).
	
	\bibitem{Soltani:16}
	\bibinfo{author}{Soltani, M.} \emph{et~al.}
	\newblock \bibinfo{title}{Ultrahigh q whispering gallery mode electro-optic
		resonators on a silicon photonic chip}.
	\newblock \emph{\bibinfo{journal}{Opt. Lett.}} \textbf{\bibinfo{volume}{41}},
	\bibinfo{pages}{4375--4378} (\bibinfo{year}{2016}).
	
	\bibitem{doi.org/10.3390/molecules25173925}
	\bibinfo{author}{Lu, Y.}, \bibinfo{author}{Johnston, B.},
	\bibinfo{author}{Dekker, P.}, \bibinfo{author}{Withford, M.~J.} \&
	\bibinfo{author}{Dawes, J.~M.}
	\newblock \bibinfo{title}{Channel waveguides in lithium niobate and lithium
		tantalate}.
	\newblock \emph{\bibinfo{journal}{Molecules}} \textbf{\bibinfo{volume}{25}},
	\bibinfo{pages}{3925} (\bibinfo{year}{2020}).
	
	\bibitem{yan2020high}
	\bibinfo{author}{Yan, X.} \emph{et~al.}
	\newblock \bibinfo{title}{High optical damage threshold on-chip lithium
		tantalate microdisk resonator}.
	\newblock \emph{\bibinfo{journal}{Optics Letters}}
	\textbf{\bibinfo{volume}{45}}, \bibinfo{pages}{4100--4103}
	(\bibinfo{year}{2020}).
	
	\bibitem{del2009frequency}
	\bibinfo{author}{Del'Haye, P.}, \bibinfo{author}{Arcizet, O.},
	\bibinfo{author}{Gorodetsky, M.~L.}, \bibinfo{author}{Holzwarth, R.} \&
	\bibinfo{author}{Kippenberg, T.~J.}
	\newblock \bibinfo{title}{Frequency comb assisted diode laser spectroscopy for
		measurement of microcavity dispersion}.
	\newblock \emph{\bibinfo{journal}{Nature Photonics}}
	\textbf{\bibinfo{volume}{3}}, \bibinfo{pages}{529--533}
	(\bibinfo{year}{2009}).
	
	\bibitem{luke2020wafer}
	\bibinfo{author}{Luke, K.} \emph{et~al.}
	\newblock \bibinfo{title}{Wafer-scale low-loss lithium niobate photonic
		integrated circuits}.
	\newblock \emph{\bibinfo{journal}{Optics Express}}
	\textbf{\bibinfo{volume}{28}}, \bibinfo{pages}{24452--24458}
	(\bibinfo{year}{2020}).
	
	\bibitem{liu2021high}
	\bibinfo{author}{Liu, J.} \emph{et~al.}
	\newblock \bibinfo{title}{High-yield, wafer-scale fabrication of ultralow-loss,
		dispersion-engineered silicon nitride photonic circuits}.
	\newblock \emph{\bibinfo{journal}{Nature Communications}}
	\textbf{\bibinfo{volume}{12}}, \bibinfo{pages}{2236} (\bibinfo{year}{2021}).
	
	\bibitem{shams2022reduced}
	\bibinfo{author}{Shams-Ansari, A.} \emph{et~al.}
	\newblock \bibinfo{title}{Reduced material loss in thin-film lithium niobate
		waveguides}.
	\newblock \emph{\bibinfo{journal}{Apl Photonics}} \textbf{\bibinfo{volume}{7}},
	\bibinfo{pages}{081301} (\bibinfo{year}{2022}).
	
	\bibitem{pan2019fundamental}
	\bibinfo{author}{Pan, A.}, \bibinfo{author}{Hu, C.}, \bibinfo{author}{Zeng, C.}
	\& \bibinfo{author}{Xia, J.}
	\newblock \bibinfo{title}{Fundamental mode hybridization in a thin film lithium
		niobate ridge waveguide}.
	\newblock \emph{\bibinfo{journal}{Optics express}}
	\textbf{\bibinfo{volume}{27}}, \bibinfo{pages}{35659--35669}
	(\bibinfo{year}{2019}).
	
	\bibitem{wang2020polarization}
	\bibinfo{author}{Wang, J.}, \bibinfo{author}{Chen, P.}, \bibinfo{author}{Dai,
		D.} \& \bibinfo{author}{Liu, L.}
	\newblock \bibinfo{title}{Polarization coupling of $ x $-cut thin film lithium
		niobate based waveguides}.
	\newblock \emph{\bibinfo{journal}{IEEE Photonics Journal}}
	\textbf{\bibinfo{volume}{12}}, \bibinfo{pages}{1--10} (\bibinfo{year}{2020}).
	
	\bibitem{wooten2000review}
	\bibinfo{author}{Wooten, E.~L.} \emph{et~al.}
	\newblock \bibinfo{title}{A review of lithium niobate modulators for
		fiber-optic communications systems}.
	\newblock \emph{\bibinfo{journal}{IEEE Journal of selected topics in Quantum
			Electronics}} \textbf{\bibinfo{volume}{6}}, \bibinfo{pages}{69--82}
	(\bibinfo{year}{2000}).
	
	\bibitem{herr2014temporal}
	\bibinfo{author}{Herr, T.} \emph{et~al.}
	\newblock \bibinfo{title}{Temporal solitons in optical microresonators}.
	\newblock \emph{\bibinfo{journal}{Nature Photonics}}
	\textbf{\bibinfo{volume}{8}}, \bibinfo{pages}{145--152}
	(\bibinfo{year}{2014}).
	
	\bibitem{kippenberg2018dissipative}
	\bibinfo{author}{Kippenberg, T.~J.}, \bibinfo{author}{Gaeta, A.~L.},
	\bibinfo{author}{Lipson, M.} \& \bibinfo{author}{Gorodetsky, M.~L.}
	\newblock \bibinfo{title}{Dissipative kerr solitons in optical
		microresonators}.
	\newblock \emph{\bibinfo{journal}{Science}} \textbf{\bibinfo{volume}{361}},
	\bibinfo{pages}{eaan8083} (\bibinfo{year}{2018}).
	
	\bibitem{stone2018thermal}
	\bibinfo{author}{Stone, J.~R.} \emph{et~al.}
	\newblock \bibinfo{title}{Thermal and nonlinear dissipative-soliton dynamics in
		kerr-microresonator frequency combs}.
	\newblock \emph{\bibinfo{journal}{Physical review letters}}
	\textbf{\bibinfo{volume}{121}}, \bibinfo{pages}{063902}
	(\bibinfo{year}{2018}).
	
	\bibitem{liu2020photonic}
	\bibinfo{author}{Liu, J.} \emph{et~al.}
	\newblock \bibinfo{title}{Photonic microwave generation in the x-and k-band
		using integrated soliton microcombs}.
	\newblock \emph{\bibinfo{journal}{Nature Photonics}}
	\textbf{\bibinfo{volume}{14}}, \bibinfo{pages}{486--491}
	(\bibinfo{year}{2020}).
	
	\bibitem{He2023}
	\bibinfo{author}{He, Y.} \emph{et~al.}
	\newblock \bibinfo{title}{High-speed tunable microwave-rate soliton microcomb}.
	\newblock \emph{\bibinfo{journal}{Nature Communications}}
	\textbf{\bibinfo{volume}{14}}, \bibinfo{pages}{3467} (\bibinfo{year}{2023}).
	
	\bibitem{zhao2020high}
	\bibinfo{author}{Zhao, J.}, \bibinfo{author}{Ma, C.},
	\bibinfo{author}{R{\"u}sing, M.} \& \bibinfo{author}{Mookherjea, S.}
	\newblock \bibinfo{title}{High quality entangled photon pair generation in
		periodically poled thin-film lithium niobate waveguides}.
	\newblock \emph{\bibinfo{journal}{Physical review letters}}
	\textbf{\bibinfo{volume}{124}}, \bibinfo{pages}{163603}
	(\bibinfo{year}{2020}).
	
	\bibitem{nehra2022few}
	\bibinfo{author}{Nehra, R.} \emph{et~al.}
	\newblock \bibinfo{title}{Few-cycle vacuum squeezing in nanophotonics}.
	\newblock \emph{\bibinfo{journal}{Science}} \textbf{\bibinfo{volume}{377}},
	\bibinfo{pages}{1333--1337} (\bibinfo{year}{2022}).
	
	\bibitem{he2019self}
	\bibinfo{author}{He, Y.} \emph{et~al.}
	\newblock \bibinfo{title}{Self-starting bi-chromatic linbo 3 soliton
		microcomb}.
	\newblock \emph{\bibinfo{journal}{Optica}} \textbf{\bibinfo{volume}{6}},
	\bibinfo{pages}{1138--1144} (\bibinfo{year}{2019}).
	
	\bibitem{gong2022monolithic}
	\bibinfo{author}{Gong, Z.}, \bibinfo{author}{Shen, M.}, \bibinfo{author}{Lu,
		J.}, \bibinfo{author}{Surya, J.~B.} \& \bibinfo{author}{Tang, H.~X.}
	\newblock \bibinfo{title}{Monolithic kerr and electro-optic hybrid microcombs}.
	\newblock \emph{\bibinfo{journal}{Optica}} \textbf{\bibinfo{volume}{9}},
	\bibinfo{pages}{1060--1065} (\bibinfo{year}{2022}).
	
	\bibitem{youssefi2021cryogenic}
	\bibinfo{author}{Youssefi, A.} \emph{et~al.}
	\newblock \bibinfo{title}{A cryogenic electro-optic interconnect for
		superconducting devices}.
	\newblock \emph{\bibinfo{journal}{Nature Electronics}}
	\textbf{\bibinfo{volume}{4}}, \bibinfo{pages}{326--332}
	(\bibinfo{year}{2021}).
	
	\bibitem{jacob2004temperature}
	\bibinfo{author}{Jacob, M.~V.} \emph{et~al.}
	\newblock \bibinfo{title}{Temperature dependence of permittivity and loss
		tangent of lithium tantalate at microwave frequencies}.
	\newblock \emph{\bibinfo{journal}{IEEE transactions on microwave theory and
			techniques}} \textbf{\bibinfo{volume}{52}}, \bibinfo{pages}{536--541}
	(\bibinfo{year}{2004}).
	
	\bibitem{yang2007characteristics}
	\bibinfo{author}{Yang, R.-Y.}, \bibinfo{author}{Su, Y.-K.},
	\bibinfo{author}{Weng, M.-H.}, \bibinfo{author}{Hung, C.-Y.} \&
	\bibinfo{author}{Wu, H.-W.}
	\newblock \bibinfo{title}{Characteristics of coplanar waveguide on lithium
		niobate crystals as a microwave substrate}.
	\newblock \emph{\bibinfo{journal}{Journal of applied physics}}
	\textbf{\bibinfo{volume}{101}}, \bibinfo{pages}{014101}
	(\bibinfo{year}{2007}).
	
\end{thebibliography}

\end{document}